\documentclass[pdflatex,sn-aps]{sn-jnl}

\usepackage{graphicx}%
\usepackage{multirow}%
\usepackage{amsmath,amssymb,amsfonts}%
\usepackage{amsthm}%
\usepackage{mathrsfs}%
\usepackage[title]{appendix}%
\usepackage{xcolor}%
\usepackage{textcomp}%
\usepackage{manyfoot}%
\usepackage{booktabs}%
\usepackage{algorithm}%
\usepackage{algorithmicx}%
\usepackage{algpseudocode}%
\usepackage{listings}%
\usepackage[dvipsnames]{xcolor}

\theoremstyle{thmstyleone}%
\theoremstyle{thmstyletwo}%

\theoremstyle{thmstylethree}%

\begin{document}

\title[Affective and Conversational Predictors in HRI Engagement]{Affective and Conversational Predictors of Re-Engagement in Human–Robot Interactions: A Student-Centered Study with A Humanoid Social Robot}


\author*[1]{\fnm{Hangyeol} \sur{Kang}}\email{hangyeol.kang@unige.ch}

\author[2]{\fnm{Thiago} \sur{Freitas}}\email{thiago.freitas@miralab.ch}

\author[3]{\fnm{Maher} \sur{Ben Moussa}}\email{maher.benmoussa@unige.ch}

\author[2]{\fnm{Nadia} \sur{Thalmann}}\email{nadia.thalmann@unige.ch}

\affil*[1]{\orgdiv{Department of Computer Science}, \orgname{University of Geneva}, \orgaddress{\street{7 Route de Drize}, \city{Carouge}, \postcode{1227}, \state{Geneva}, \country{Switzerland}}}

\affil[2]{\orgdiv{MIRALab}, \orgname{University of Geneva}, \orgaddress{\street{7 Route de Drize}, \city{Carouge}, \postcode{1227}, \state{Geneva}, \country{Switzerland}}}

\affil[3]{\orgdiv{Centre Universitaire d'Informatique}, \orgname{University of Geneva}, \orgaddress{\street{7 Route de Drize}, \city{Carouge}, \postcode{1227}, \state{Geneva}, \country{Switzerland}}}


\abstract{Humanoid social robots are increasingly present in daily life, making sustained user engagement a critical factor for their effectiveness and acceptance. While prior work has often examined affective evaluations or anthropomorphic design, less is known about the relative influence of dynamic conversational qualities and perceived robot characteristics in determining a user's intention to re-engage with Large Language Model (LLM)-driven social robots. In this study, 68 participants interacted in open-ended conversations with the Nadine humanoid social robot, completing pre- and post-interaction surveys to assess changes in robot perception, conversational quality, and intention to re-engage. The results showed that verbal interaction significantly improved the robot's perceived characteristics, with statistically significant increases in pleasantness ($p<.0001$) and approachability ($p<.0001$), and a reduction in creepiness ($p<.001$). However, these affective changes were not strong and unique predictors of users' intention to re-engage in a multiple regression model. Instead, participants' perceptions of the interestingness ($\beta=0.60$, $p<.001$) and naturalness ($\beta=0.31$, $p=0.015$) of the robot's conversation emerged as the most significant and robust independent predictors of intention to re-engage. Overall, the results highlight that conversational quality, specifically perceived interestingness and naturalness, is the dominant driver of re-engagement, indicating that LLM-driven robot design should prioritize engaging, natural dialogue over affective impression management or anthropomorphic cues.}

\keywords{Social robots, Human-robot interaction, User study, User re-engagement, Affective evaluation, Conversational quality}



\maketitle

\begin{center}
\small
This manuscript has been accepted for publication in \textit{International Journal of Social Robotics}. The final authenticated version will be available from the publisher.
\end{center}

\section{Introduction}
\label{sec:1}
Social robots have emerged as transformative tools in various sectors, including healthcare, education, and hospitality, due to their ability to interact with humans in socially meaningful ways~\cite{vishwakarma2024adoption}. These robots are designed not only to perform tasks but also to provide emotional support, foster social engagement, and enhance user experiences. For example, in healthcare, social robots assist elderly patients by offering companionship~\cite{mishra2022nadine} and supporting mental well-being~\cite{laban2024building}, while in education, they serve as tutors or co-learners to promote cognitive and social development~\cite{donnermann2022social}. As these robots move beyond experimental prototypes and begin to interact with the general public, their success depends largely on their ability to foster meaningful and sustained engagement with users.

User engagement, particularly the willingness to continue interacting with a robot, is widely recognized as a critical factor for the effectiveness and acceptance of social robots~\cite{chen2025impact,de2017they}. Without sustained interaction, even the most advanced robots may fail to deliver their intended benefits, whether in education, customer service, therapy, or companionship. Therefore, understanding the psychological and communicative factors that drive users to initiate, maintain, or discontinue interactions with humanoid robots has become a central research focus in the field of human-robot interaction (HRI)~\cite{kaduk2024emotional,irfan2024human}.

Engagement in HRI is a multifaceted construct shaped by a range of psychological, social, and design-related factors~\cite{sorrentino2024definition}. Previous research has identified several key elements that contribute to users' willingness to interact with social robots, including emotional responses, the physical and behavioral design of the robot, and the degree of anthropomorphism in its appearance and behavior~\cite{chen2025impact,novikova2017sympathy}. Factors such as perceived safety, trustworthiness, and comfort have also been shown to affect the quality and duration of HRI~\cite{akalin2022you,redondo2024comfortability,miller2021more}.

Among these, affect plays a central role in determining engagement outcomes~\cite{sorrentino2024definition}. Positive emotional responses, such as increased pleasantness or reduced anxiety, have been linked to greater acceptance and longer interaction times, while negative emotions can quickly disrupt or end the engagement~\cite{smith2020positive,bakan2025exploring}. Emotional reactions are often influenced by both the robot's appearance and its communicative behaviors, with subtle differences in design or expression potentially leading to significant shifts in user perceptions~\cite{lu2025effects,kaduk2024emotional}.

In addition to affect, the perceived quality of the robot's communication abilities is increasingly recognized as a major determinant of user engagement~\cite{park2024av,leoste2024nonverbal}. Attributes such as the naturalness, human-likeness, and interestingness of verbal exchanges can shape how users evaluate the robot's social competence and decide whether to continue the interaction~\cite{van2020human}. Well-designed communication cues can make interactions feel smoother and more enjoyable, enhancing the overall user experience and supporting the formation of ongoing social relationships with robots~\cite{irfan2024human,reimann2025can}.

Despite the recognition that affective factors play a significant role in HRI, much of the existing research has treated emotion primarily as an outcome variable, as reviewed by Stock et al.~\cite{stock2022survey}. However, comparatively few studies have explored how changes in users' perceived robot characteristics~\cite{li2023effects,yang2025study}, particularly those elicited by the robot's verbal communication, translate into concrete intention to re-engage. The gap is important, as understanding the dynamic relationship between these perceptual shifts and sustained interaction can offer practical insights for designing socially engaging robots. 

Similarly, while conversational qualities like naturalness, human-likeness, and interestingness are frequently assessed in HRI studies~\cite{irfan2024human,van2020human}, most research has focused on subjective user ratings of these attributes rather than their direct predictive impact on users' motivation to persist in the interaction. The extent to which improvements in conversational quality foster continued engagement remains underexplored, limiting our understanding of what truly drives sustained interaction with social robots.

Additionally, a considerable portion of the literature has emphasized the role of anthropomorphism as a central strategy for enhancing engagement~\cite{shum2024so,bakan2025exploring}. Yet, evidence regarding its effectiveness remains mixed, with some studies reporting positive effects~\cite{kim2019eliza} while others find negative consequences~\cite{bakan2025exploring}. This suggests a need to move beyond anthropomorphism as the primary lens and to investigate which specific psychological factors (perceived robot characteristics) and communicative factors (conversational quality) serve as strong independent predictors of users' re-engagement.

To address these gaps, the present study focuses on three central research questions:
\begin{itemize}
    \item RQ1: How do participants' perceived robot characteristics (pleasantness, creepiness, and approachability) change after a one-on-one verbal interaction with the social robot?
    \item RQ2: How do changes in perceived robot characteristics (pleasantness, creepiness, and approachability) predict participants' intention to re-engage with the robot? 
    \item RQ3: How do perceived conversational qualities (naturalness, human-likeness, interestingness) influence participants' intention to re-engage with the robot?
\end{itemize}

By directly examining these questions, this study aims to move beyond static measures of user experience and instead investigate the dynamic psychological processes that drive engagement with social robots. There is a clear need for empirical evidence that links changes in robot perception and perceptions of conversational quality with concrete intention to re-engage. Such evidence is crucial for developing a more comprehensive understanding of what fosters meaningful and lasting engagement in human-robot interaction.

This study makes three main contributions to the field of HRI. 

\begin{figure}[h]
    \centering
    \includegraphics[width=0.8\linewidth]{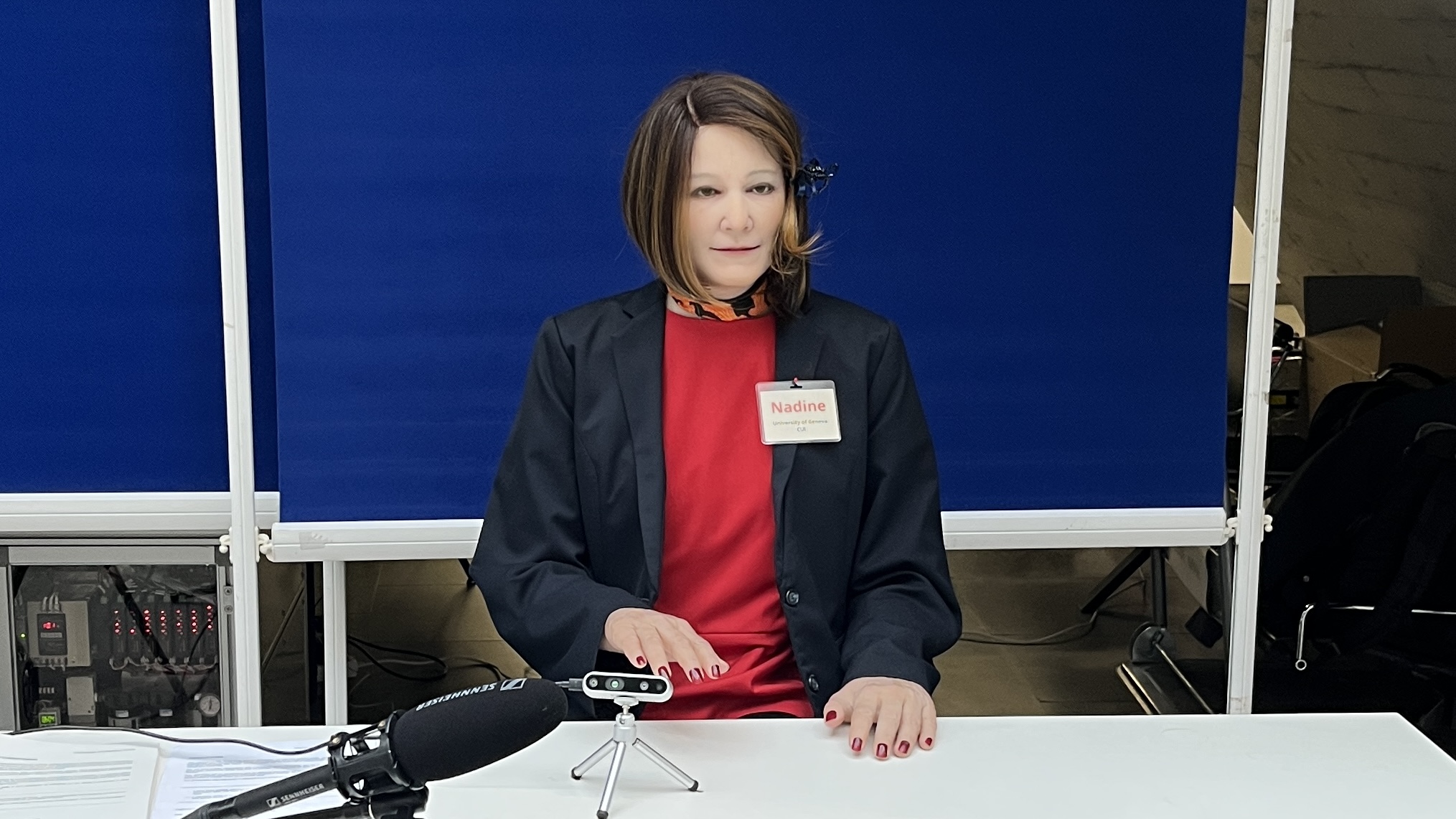}
    \caption{The Nadine humanoid social robot used in this study during the one-on-one interaction setup.}
    \label{fig:NadineFront}
\end{figure}

First, it provides quantitative empirical evidence on the relative predictive strength of affective factors (operationalized as changes in perceived robot characteristics) and conversational qualities (naturalness, human-likeness, and interestingness) on users' intention to re-engage with the LLM-driven humanoid social robot Nadine (Figure~\ref{fig:NadineFront}). Most prior HRI studies evaluate users' affective impressions at a single point rather than examining how shifts in these perceptions relate to subsequent engagement behavior~\cite{reeves2020social,naneva2020systematic}. This direct comparison addresses an important gap in existing HRI literature.

Second, the findings offer practical insights for the design and development of socially engaging robots. By identifying which aspects of verbal communication (e.g., interestingness) are most influential in shaping user motivation, this study can inform future design strategies that prioritize advanced, high-quality dialogue systems over merely optimizing robot appearance or affective evaluations.

Finally, our findings highlight that the authenticity and quality of the robot’s conversational abilities, specifically, the interestingness and naturalness of its dialogue, play a more decisive role in sustaining user engagement than the perceived human-likeness of its conversational style alone. This distinction challenges the traditional emphasis on anthropomorphic design and underscores the need to prioritize engaging and natural communication when aiming to foster meaningful and lasting human-robot relationships.

The remainder of this paper is structured as follows: Section \ref{sec:2} reviews related work on emotion and HRI and communication qualities and HRI. Section \ref{sec:3} outlines our methodology, while Section \ref{sec:4} presents the results. Section \ref{sec:5} discusses implications for design and future research directions. Finally, Section \ref{sec:6} concludes the study.

\section{Related Work}
\label{sec:2}
\subsection{Affective Responses and Perceived Robot Characteristics in HRI}
Affective responses are widely recognized as a core dimension of user evaluation in human-robot interaction~\cite{stock2022survey}. Research across long-term and short-term HRI consistently shows that users form emotional impressions, such as comfort, warmth, or unease, that influence how they perceive a robot's social qualities and whether they are willing to interact with it~\cite{leite2013social,abdi2022study}. These affective judgments function as fundamental appraisal mechanisms through which users assess whether a robot feels comfortable, safe, and socially appropriate~\cite{redondo2024comfortability,rubagotti2022perceived,stapels2023never}. In line with this literature, the present study examines three key perceptual characteristics, specifically pleasantness, creepiness, and approachability, to capture users' affective impressions of the robot.

Pleasantness and comfort have been repeatedly identified as predictors of user acceptance. Studies examining empathic and socially responsive robots demonstrate that positive affective evaluations lead to higher satisfaction and more favorable impressions of a robot's sociability and competence~\cite{birnbaum2016robots,abdollahi2022artificial}. Empirical work in service and assistive contexts likewise shows that pleasant affect during interaction is associated with higher perceived quality of the interaction and more positive attitudes toward the robot~\cite{choi2020service,fang2024hotel}. These findings support the treatment of pleasantness as a meaningful and widely adopted indicator of users' affective assessment.

Conversely, negative affective responses, such as creepiness, eeriness, or discomfort, have been linked to interaction avoidance. The uncanny valley framework~\cite{mori1970uncanny} and subsequent empirical studies~\cite{grazzini2023dashed} show that robots that fall into this perceptual gap elicit discomfort and reduced willingness to engage. Standardized instruments such as the Robotic Social Attributes Scale (RoSAS) further codify discomfort as a core perceptual dimension in robot evaluation~\cite{carpinella2017robotic}, while newer work applying the Perceived Creepiness of Technology Scale demonstrates that creepiness is a measurable and relevant construct in modern HRI~\cite{wozniak2021creepy}. Together, these studies establish creepiness as an affective response with clear behavioral implications.

A third construct relevant to affective evaluation is approachability, which reflects perceptions of the robot's social openness and accessibility. Research shows that users' comfort in approaching or being approached by a robot is shaped by their affective judgments of its behavior and appearance~\cite{walters2009empirical,neef2023appropriate}. More recent findings indicate that subtle social behaviors, such as responding to joint attention cues, can increase perceived social attributes related to approachability~\cite{garcia2024analyzing}. In decision-making contexts, approachability has also been shown to mediate whether users choose to socially engage with a robot~\cite{tulk2018trust}, highlighting its relevance as a predictor of behavioral intentions.

Despite these insights, prior literature leaves two important gaps. First, most studies examine affective impressions at a single point in time~\cite{reeves2020social,naneva2020systematic} rather than assessing how pleasantness, creepiness, or approachability change after an actual conversational exchange. Second, while affective evaluations are known to shape attitudes and initial acceptance~\cite{stock2022survey}, comparatively little work has examined whether changes in these perceptual characteristics predict users' intention to re-engage with a humanoid robot~\cite{yang2025study,premathilake2024users}. Addressing these gaps is essential for understanding how dynamic affective processes contribute to sustained engagement in HRI, a focus that directly motivates the approach taken in this study.

\subsection{Communicational Qualities in HRI}
Verbal communication is a central mechanism through which social robots establish rapport, convey intentions, and sustain meaningful interaction with users. Research shows that conversational behaviors, particularly timing, responsiveness, and grounding, substantially influence how users perceive a robot's social competence and the overall quality of the interaction experience~\cite{skantze2021turn}. As robots increasingly rely on advanced language models to participate in open-domain dialogue, there is growing interest in identifying which aspects of conversational performance matter most to users~\cite{sakamoto2025effectiveness}. Recent frameworks such as HRI CUES~\cite{irfan2024human} provide systematic constructs for evaluating conversational qualities in HRI, and emerging work on LLM-driven dialogue systems highlights the need to assess user-centered conversational attributes in these more open-ended settings~\cite{marcinek2025role}.

Building on this foundation, three conversational qualities have been particularly emphasized in HRI research: naturalness~\cite{miller2025timing,lin2022duplex}, human-likeness~\cite{van2020human,sirithunge2021evaluation,jang2024beyond}, and interestingness~\cite{velentza2021one,irfan2024human,marcinek2025role}. Naturalness encompasses features such as prosodic appropriateness, response timing, and smooth coordination of turns, all of which shape users’ comfort and perceptions of a robot’s social competence. Prior work shows that delays, mistimed responses, or violations of conversational rhythm reduce perceived naturalness and negatively affect judgments of a robot’s responsiveness, intelligence, and social fluency~\cite{miller2025timing,lin2022duplex}.

Human-likeness, often expressed through prosody, voice, and emotional tone, can enhance relational perceptions such as trust, rapport, and social presence~\cite{kluber2025affect}. However, meta-analytic and review evidence indicate that these effects vary across contexts and that human-like behaviors do not reliably translate into deeper or sustained engagement~\cite{van2020human,sirithunge2021evaluation}. Even when human-like cues increase perceived friendliness or sociability, they may fail to produce consistent long-term interaction benefits if they are not aligned with user expectations or situational context~\cite{jang2024beyond}.

Interestingness, though less frequently operationalized in HRI, has emerged as an important contributor to users’ affective experience. Studies of robot storytelling demonstrate that expressive communication styles and engaging presentation reliably increase user enjoyment and shape preferences for future collaboration~\cite{velentza2021one}. Likewise, recent frameworks for assessing conversational enjoyment identify enjoyment, shaped by interest, stimulation, and engagement, as a central dimension of users’ subjective experience during robot interaction~\cite{irfan2024human,marcinek2025role}.

Despite these advances, existing research has several limitations that constrain our understanding of how conversational qualities influence sustained engagement. Most studies examine naturalness, human-likeness, or interestingness independently, without comparing their relative contributions to user experience~\cite{miller2025timing,jang2024beyond}. Moreover, prior work typically focuses on momentary enjoyment or satisfaction, rather than on users' intention to re-engage, a critical behavioral indicator of sustained interaction~\cite{velentza2021one,marcinek2025role}. Finally, little empirical evidence addresses how conversational qualities predict future engagement. These gaps make it unclear which aspects of conversation matter most for motivating continued interaction with social robots.

The present study addresses these limitations by jointly examining perceived naturalness, human-likeness, and interestingness as predictors of users' intention to re-engage, and by evaluating their predictive strength following an LLM-driven interaction. \\

\section{Method}
\label{sec:3}
\subsection{Participants}
A total of 68 participants were randomly recruited from the University of Geneva to take part in this study. Participants ranged in age from 18 to 34 years (M=23, SD=3.61), with 46 identifying as female and 22 as male. The sample included individuals from a variety of academic disciplines, with primary representation from law, economics, management, psychology, and education. Fluency in either English or French was required for participation, as these were the languages used in the robot interactions. Eleven participants interacted with the robot in English, while the remaining fifty-seven interacted in French. Participants reflected diverse geographic backgrounds, though the majority were residents of Switzerland. All participants provided informed consent for their participation and for the use of their data in scientific research. The study received full ethical approval from the University of Geneva's Committee for Ethical Research (CUREG-2024-10-109).

\subsection{Experimental Setup and Procedure}
\begin{figure}[h]
    \centering
    \includegraphics[width=\linewidth]{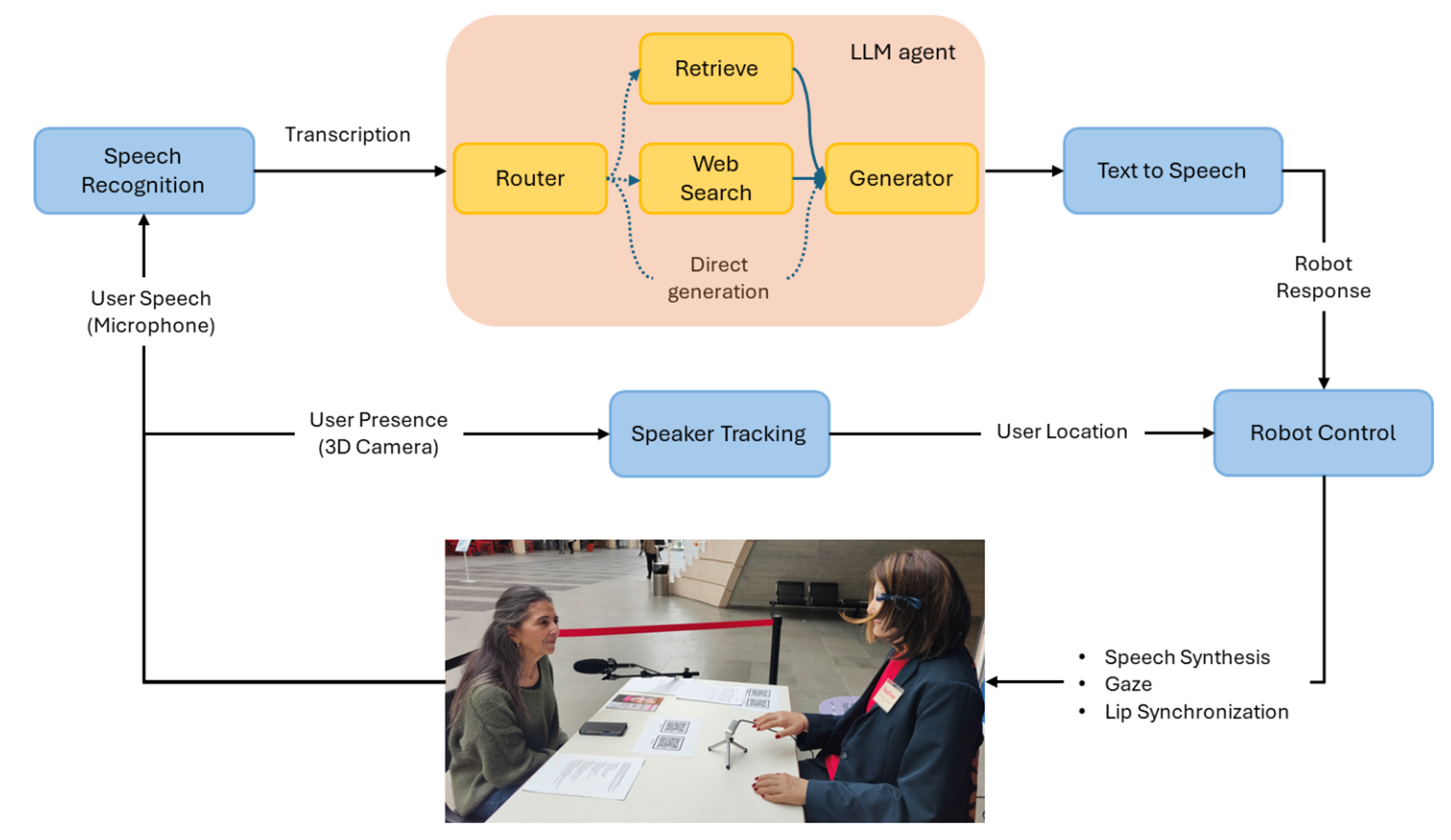}
    \caption{Nadine system architecture.}
    \label{fig:NadinePlatform}
\end{figure}

In this study, the robot was placed in the lobby of the University of Geneva (Uni Mail), chosen for its high accessibility and visibility to maximize participant recruitment. The location allowed participants to approach the robot naturally while providing a semi-controlled environment for the experiment. To ensure a smooth and controlled experiment, interactions between the robot and participants were conducted in a one-on-one manner. The interaction space was delineated from the surrounding environment using retractable barriers, as illustrated in Figure \ref{fig:NadinePlatform}, creating a clearly defined interaction zone that reduced external distractions.

Before interacting with the robot, participants completed a pre-interaction questionnaire. This questionnaire gathered demographic information, such as age, gender, and prior experience with robots, and assessed participants' initial perceptions of the robot based solely on its appearance and presence. These baseline responses served as a reference to measure changes in perception following the interaction.

Participants were invited to engage in open-ended conversations with the robot, during which they could freely explore topics of interest and ask questions without predetermined constraints. Participants retained full autonomy over the interaction, with the ability to conclude their conversation at their discretion. Experiment completion times ranged from 2 to 18 minutes (M=6.15, SD=3.46).

After the interaction, participants completed a post-interaction questionnaire designed to capture their impressions of the robot across several dimensions. All items were rated on a five-point Likert scale (1 = not at all, 5 = very much). Affective responses (creepiness, pleasantness, and approachability) were measured both before and after the interaction to capture any shifts resulting from direct engagement with the robot. Additionally, immediately following the interaction, participants rated the perceived qualities of the robot's conversation, specifically its naturalness, human-likeness, and interestingness. Finally, participants indicated their intention to re-engage with the robot, which served as the primary outcome variable for this study. This design allowed us to examine how changes in affective responses and post-interaction perceptions of conversational qualities predicted participants' intention to sustain engagement with the robot.

\subsection{Evaluation Metrics}
In order to assess participants' experience and engagement with the humanoid social robot, we selected a set of evaluation metrics grounded in prior research on HRI. These metrics reflect both affective responses and perceived conversational qualities, as well as their link to behavioral intentions.

We measured pleasantness, creepiness, and approachability as key affective dimensions relevant to initial and ongoing interactions with social robots. The concept of `creepiness' is particularly salient when evaluating user comfort and acceptance of social robots, as it addresses feeling of unease that can emerge from human-like technologies~\cite{wozniak2021creepy}. Pleasantness and approachability reflect the positive valence and social warmth that robots can elicit~\cite{kuhne2020human,tulk2018trust}. Approachability was assessed to reflect participants' perceptions of how easy, comfortable, and socially inviting the robot seemed.

For conversational qualities, naturalness and human-likeness were selected as they represent core attributes users seek in social interactions with robots~\cite{miller2025timing,jang2024beyond}. Interestingness, as measured by recent frameworks like the HRI CUES (Conversational User Enjoyment Scale)~\cite{irfan2024human}, is closely associated with conversational enjoyment and long-term user engagement.

Finally, intention to re-engage serves as a primary outcome variable because it directly reflects users' behavioral intentions and is a well-established indicator of social robot acceptance and the likelihood of repeated engagement. Both the HRI CUES and SERVBOT~\cite{kharub2021perceived} models highlight the importance of affective and conversational experiences in predicting users' willingness to sustain interaction with robots.

Collectively, these variables enable a comprehensive assessment of the affective, conversational, and behavioral factors underlying successful HRI.

\subsection{Humanoid Social Robot, Nadine}
The Nadine robot’s full system architecture incorporates a comprehensive set of cognitive, memory, and affective modules designed to support rich, human-like social interaction~\cite{kang2024nadine}. These include a long-term memory module that stores user-specific episodic information across sessions, a RAG-based knowledge retrieval system, and a multi-layer affective system that models emotions, mood, and personality to generate emotionally congruent behavior. While these modules are essential for achieving personalized and affectively adaptive interaction, they also introduce behavioral variability driven by the robot’s accumulated memory states or emotional dynamics, rather than by the verbal interaction alone. Because the present study aims to isolate how Nadine’s verbal communication alone influences users’ affective evaluations and re-engagement intentions, we employed a streamlined version of the platform (Fig.~\ref{fig:NadinePlatform}) that excludes long-term memory and affective modules. This controlled configuration ensures that participants’ perceptions reflect only the robot’s immediate conversational behavior, not personalization or affect-driven adaptation.

The LLM-agent implemented in this study is built on LangGraph\footnote[1]{\url{https://www.langchain.com/langgraph}}, a framework for implementing multi-step and multi-module agent workflows. As shown in Fig.~\ref{fig:NadinePlatform}, the agent comprises four primary modules: router, retriever, web search, and generator. The router serves as the central decision-making unit, selecting the optimal processing pathway for each user query. To support real-time interaction, we employed OpenAI's gpt-4o-mini-2024-07-18\footnote[2]{\url{https://platform.openai.com/docs/models/gpt-4o-mini}}, chosen for its lower computational cost and faster inference compared with larger models. Across the study, the system produced a mean generation latency of 2.514 s and a variance of 0.296 s, which enabled smooth, continuous interaction during the experiment.

The generator module operated with a fixed system prompt designed to maintain Nadine's human-like persona and situational grounding. The system prompt instructed the model to speak in the first person as Nadine, adopt human-like preferences, and keep responses concise. It also included a short physical context description and dynamically injected retrieved knowledge when available. A simplified version of the prompt is provided below: `` Your name is Nadine and you are a humanoid social robot... You are in the University of Geneva talking to a person sitting in front of you... Always speak in the first person... Context: {context}. ''

For queries requiring knowledge of the robot's background or prior experience, the router directs requests to the retriever module, which accesses relevant data from an external knowledge base implemented using Chroma vectorstore~\footnote[3]{\url{https://www.trychroma.com/}}. In contrast to the full SoR-ReAct architecture, this retriever was restricted to robot-related information only, as user-specific long-term memory was omitted for experimental consistency. The toolset was similarly simplified. The original toolset (internet search, news search, weather search, and Wikipedia) was reduced to a single web-search module using the GoogleSerperAPIWrapper~\footnote[4]{\url{https://python.langchain.com/docs/integrations/providers/google_serper/}}, which provided real-time search results when required.

If a query did not require external retrieval or live information, the router forwarded it directly to the generator module. The generator, also powered by gpt-4o-mini, produced context-sensitive responses guided by the system prompt described above. The removal of the long-term memory module also meant that user identification and personalized retrieval were not used in this study, further reducing system complexity and improving latency.

Beyond verbal generation, the system incorporated key non-verbal interaction components used in the study, namely speech synthesis, lip synchronization, and gaze control, to ensure coherent multimodal behavior during the HRI. Speech synthesis was implemented using Microsoft Azure's neural text-to-speech service (Cognitive Services Speech SDK\footnote[5]{\url{https://github.com/Azure-Samples/cognitive-services-speech-sdk}}), which returns both high-quality audio output and time-aligned viseme events. These visemes were streamed to a customized lip-animation controller that interpolates motor movements to achieve natural lip synchronization during spoken responses. For gaze control, the robot employed a real-time RGB-D perception pipeline with a YOLOv8-based face and eye-keypoint detector, enabling estimation of the user's eye position in 3D. This estimate was used to drive the head and eye motors so that Nadine maintained gaze toward the interlocutor throughout the interaction. These non-verbal modules operate independently of the LLM but were automatically triggered whenever a generated utterance requires audio or embodied output. \\

\section{Results}
\label{sec:4}
This section presents the findings of our analyses, structured according to the three research questions. We first examine how participants' perceived robot characteristics, including pleasantness, creepiness, and approachability, changed following the verbal interaction (RQ1). We then evaluate whether these perceptual changes are associated with participants' intention to re-engage with the robot (RQ2). Finally, we assess how perceived conversational qualities, including naturalness, human-likeness, and interestingness, predict participants' intention to re-engage with the robot (RQ3).

Across these analyses, we report descriptive statistics prior to conducting correlation and regression analyses. Together, these results provide a comprehensive understanding of how changes in robot perception and evaluations of the robot's conversational abilities relate to users' intention to re-engage with a humanoid social robot.

\subsection{RQ1. Changes in Perceived Robot Characteristics}
\subsubsection{Descriptive Statistics}
To examine how participants' perceptions of the robot changed following the interaction, we first report descriptive statistics for pleasantness, creepiness, and approachability before and after the verbal interaction. Table~\ref{tab:rq1_rescriptive} summarizes the means, standard deviations, medians, and interquartile range (IQRs) for each variable. As shown, pleasantness and approachability increased on average from pre- to post-interaction, while creepiness decreased.

\begin{table}[h]
    \begin{tabular*}{\textwidth}{@{\extracolsep\fill}lcccccc}
    \toprule
    & \multicolumn{2}{@{}c@{}}{Pleasant} & \multicolumn{2}{@{}c@{}}{Creepy} & \multicolumn{2}{@{}c@{}}{Approachable} \\
    \cmidrule{2-3}\cmidrule{4-5}\cmidrule{6-7}
     & Pre & Post & Pre & Post & Pre & Post \\
    \midrule
    Mean   & 2.897 & 3.706 & 3.456 & 2.912 & 3.221 & 3.853 \\
    SD     & 0.877 & 0.956 & 0.946 & 1.209 & 1.041 & 1.047 \\
    Median & 3.0   & 4.0   & 4.0   & 3.0   & 3.0   & 4.0   \\
    IQR    & 1.0   & 1.0   & 1.0   & 2.0   & 1.0   & 2.0   \\
    \botrule
    \end{tabular*}
    \caption{Descriptive statistics for perceived robot characteristics before and after the verbal interaction. Values represent mean, standard deviation (SD), median, and interquartile range (IQR).}
    \label{tab:rq1_rescriptive}
\end{table}

\begin{figure}[h]
    \centering
    \includegraphics[width=\linewidth]{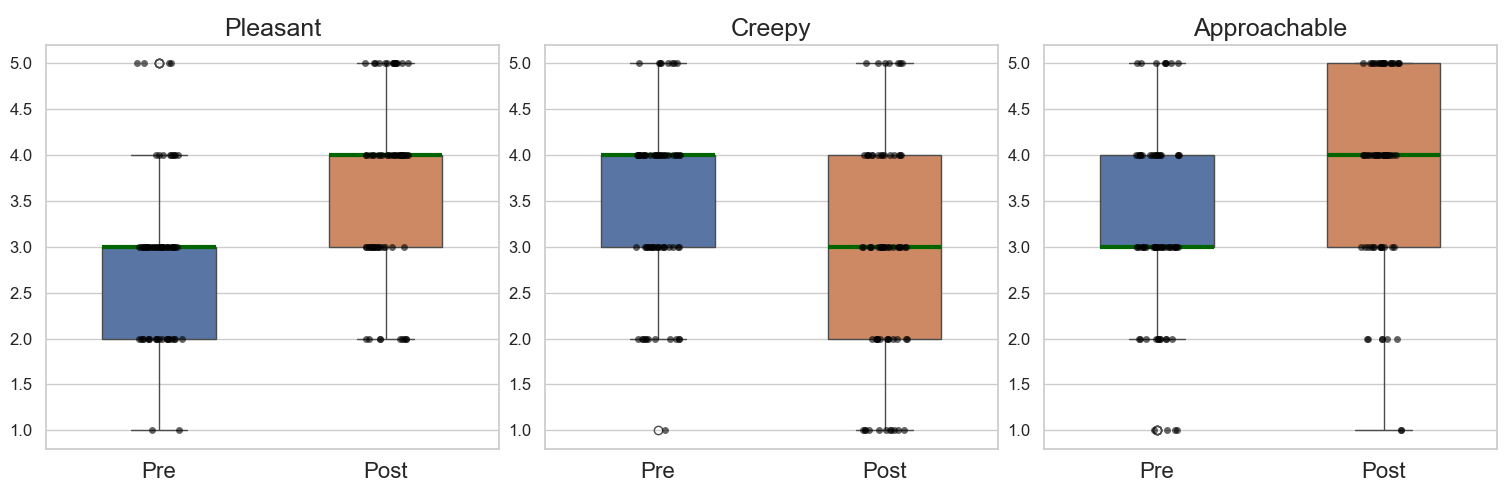}
    \caption{Boxplots showing pre- and post-interaction ratings of pleasantness, creepiness, and approachability. The boxes represent the interquartile ranges (IQRs), whiskers show the full range excluding outliers, individual points represent participant responses, and the green horizontal line indicates the median.}
    \label{fig:rq1_boxplots}
\end{figure}

Figure~\ref{fig:rq1_boxplots} visualizes the distribution of ratings using boxplots. For both pleasantness and approachability, the median values increased after the interaction, and the upper quartile shifted upward, indicating higher overall ratings among participants. In contrast, creepiness shows a lower median in the post-interaction condition. The IQR widened for creepiness and approachability after the interaction, suggesting increased variability in participants’ responses. Individual data points are plotted to illustrate the spread and density of ratings across participants.

These descriptive summaries provide an initial overview of how participants' perceived robot characteristics shifted following the verbal interaction and serve as the basis for the Wilcoxon signed-rank tests reported in the next subsection.

\subsubsection{Pre-Post Statistical Comparisons}
We conducted Wilcoxon signed-rank tests~\cite{wilcoxon1992individual, seabold2010statsmodels} to evaluate changes in participants' perceived robot characteristics following the interaction. The results revealed statistically significant changes across all three variables:

\par{\textbf{Pleasantness:}} Ratings significantly increased ($V=103.0$, $p<.0001$, $d=0.83$), indicating a large effect size and a clear improvement in how pleasant participants found the robot. A post-hoc power analysis demonstrated strong reliability ($power=1.0$), confirming the study had sufficient power to detect the effect.

\par{\textbf{Creepiness:}} Ratings significantly decreased ($V=355.5$, $p<.001$, $d=-0.39$), reflecting a small to medium effect size. Although this indicates a reduction in discomfort, the change may not be substantial for all participants. Post-hoc power analysis confirmed sufficient power ($0.9255$) to detect this effect.

\par{\textbf{Approachability:}} Ratings significantly increased ($V=244.0$, $p<.0001$, $d=0.55$), with a medium effect size and a noticeable increase in perceived approachability. Post-hoc power analysis ($power=0.9983$) further supports the robustness of these findings.

Taken together, these findings show that participants perceived the robot as more pleasant, approachable and less creepy after the verbal interaction, reflecting a broadly positive shift in perceived robot characteristics.

\subsection{RQ2. Predictive Role of Changes in Perceived Robot Characteristics}
To investigate whether changes in participants' perceptions of the robot were associated with their intention to re-engage, we examined the relationships between pre-post changes in pleasantness, creepiness, and approachability and participants' re-engagement intentions. We first computed Pearson correlations~\cite{cohen2013applied} for each perceptual change score, followed by multiple regression analyses~\cite{seabold2010statsmodels} to assess the unique contribution of each predictor when considered simultaneously.

\subsubsection{Correlation Analysis}
\label{sec:4.2.1}
\begin{figure}[h]
    \centering
    \includegraphics[width=\linewidth]{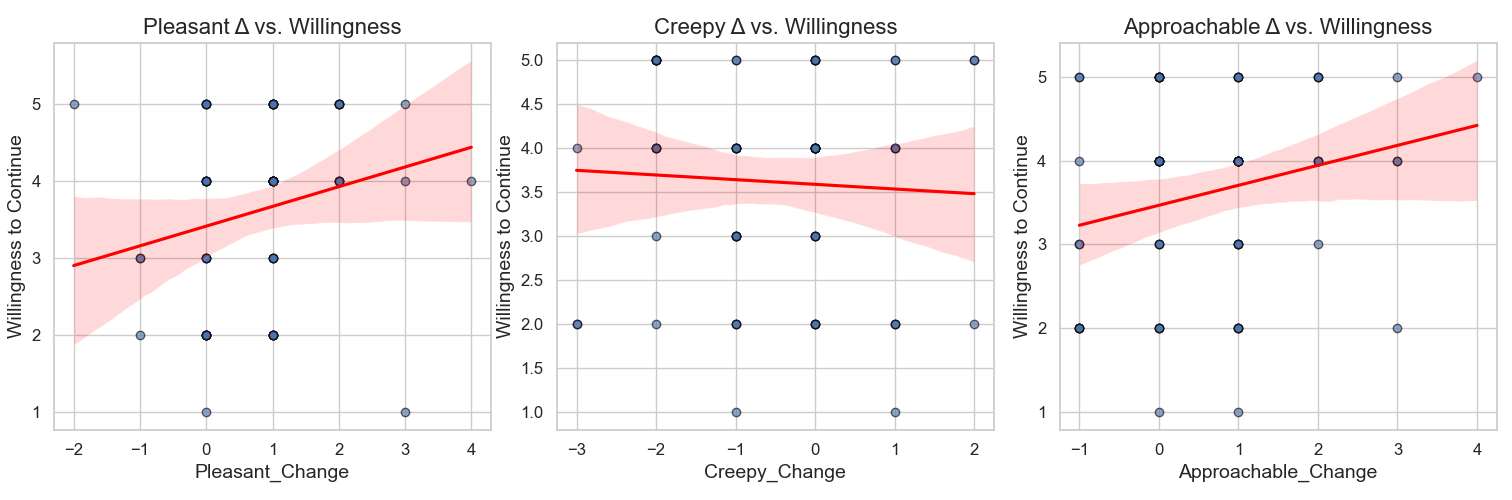}
    \caption{Scatterplots showing the relationships between changes in pleasantness, creepiness, and approachability (post minus pre) and participants’ intention to re-engage with the robot. The red line represents the fitted linear regression line, and the shaded region denotes the 95\% confidence interval.}
    \label{fig:rq2_scatter}
\end{figure}

We first examined the bivariate association between changes in perceived robot characteristics and participants' intention to re-engage with the robot. Pearson correlation coefficients, corresponding $p$-values, and 95\% confidence intervals are presented in Table~\ref{tab:rq2_corr}. Scatter plots with fitted regression lines are shown in Figure~\ref{fig:rq2_scatter}.

None of the correlations between perceptual change scores and re-engagement intention reached statistical significance. Pleasantness change ($ r=0.224$, $p=0.067$), creepiness change ($r=-0.055$, $p=0.659$), and approachability change ($r=0.230$, $p=0.059$) showed confidence intervals that included zero, indicating no evidence of reliable linear associations.

Given the absence of significant bivariate relationships, no further conclusions are drawn from these results. We proceeded to a multiple regression analysis to examine the unique contributions of each predictor when considered simultaneously.

\begin{table}[h]
    \begin{tabular*}{\textwidth}{@{\extracolsep\fill}lccc}
    \toprule
    Predictor & Pearson $r$ & $p$ & 95\% CI \\
    \midrule
    Pleasant Change      & 0.224 & 0.067 & [-0.015, 0.439] \\
    Creepy Change        & -0.055 & 0.659 & [-0.289, 0.186] \\
    Approachable Change  & 0.230 & 0.059 & [-0.009, 0.444] \\
    \botrule
    \end{tabular*}
    \caption{Pearson correlations between changes in perceived robot characteristics and participants’ intention to re-engage. The table reports correlation coefficients ($r$), corresponding $p$-values, and 95\% confidence intervals (CI).}
    \label{tab:rq2_corr}
\end{table}

\subsubsection{Multiple Regression Model}
\begin{figure}[h]
    \centering
    \includegraphics[width=0.9\linewidth]{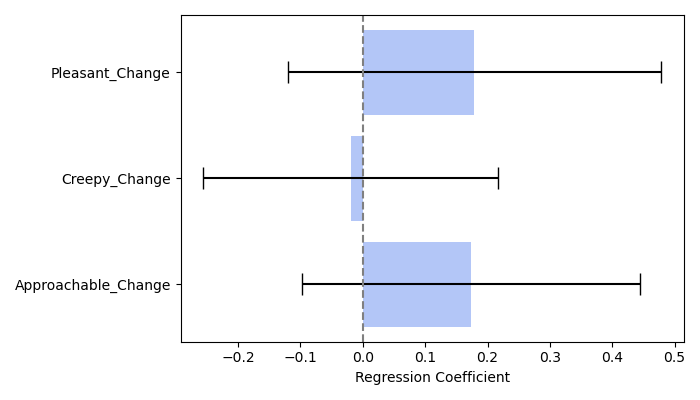}
    \caption{Coefficient plot showing the unique effects of changes in perceived robot characteristics on participants' intention to re-engage, based on a multiple linear regression model. Each bar represents the unstandardized regression coefficient, and the horizontal lines indicate the 95\% confidence intervals.}
    \label{fig:rq2_coefpplot}
\end{figure}

To assess the unique contributions of each perceptual change variable to participants' intention to re-engage, we conducted a multiple linear regression including pleasantness change, creepiness change, and approachability change as simultaneous predictors. Figure~\ref{fig:rq2_coefpplot} displays the unstandardized regression coefficients with 95\% confidence intervals, and Table~\ref{tab:rq2_regression} reports the full model estimates.

\begin{table}[h]
    \begin{tabular*}{\textwidth}{@{\extracolsep\fill}lcccccc}
    \toprule
    Variable & coef. & std err & $t$ & $p$ & std.\ coef. & 95\% CI \\
    \midrule
    Pleasant Change      & 0.179 & 0.150 & 1.198 & 0.236 & 0.157 & [-0.120, 0.478] \\
    Creepy Change        & -0.019 & 0.118 & -0.163 & 0.871 & -0.020 & [-0.255, 0.217] \\
    Approachable Change  & 0.173 & 0.135 & 1.277 & 0.206 & 0.167 & [-0.098, 0.444] \\
    \botrule
    \end{tabular*}
    \caption{Multiple linear regression results predicting intention to re-engage from changes in perceived robot characteristics. Unstandardized coefficients (coef.), standard errors (std err), $t$-values, $p$-values, standardized coefficients (std. coef.), and 95\% confidence intervals (CI) are reported.}
    \label{tab:rq2_regression}
\end{table}

None of the predictors emerged as statistically significant when entered together in the model. Pleasantness change ($p=0.236$), creepiness change ($p=0.871$), and approachability change ($p=0.206$) showed non-significant associations with re-engagement intention, and the confidence intervals for all three predictors included zero. These results indicate that, after accounting for shared variance among the perceptual change variables, no single predictor reliably explained additional variance in participants' intention to re-engage with the robot.

Consistent with the correlation analyses reported in Section~\ref{sec:4.2.1}, the regression analyses provide no evidence that changes in perceived robot characteristics independently predict intention to re-engage.

\subsection{RQ3. Predictive Role of Conversational Qualities}
In addition to changes in perceived robot characteristics, we examined whether participants' evaluations of the robot's conversational qualities, specifically naturalness, human-likeness, and interestingness, were associated with their intention to re-engage with the robot. Because these variables represent participants' immediate assessments of the robot's dialogue behavior following the interaction, we analyzed their predictive value using both correlation analyses and a multiple regression model. This approach allows us to assess the extent to which each conversational attribute uniquely contributes to re-engagement intention when considered alongside the others.

\subsubsection{Descriptive Statistics}
We first summarize participants' post-interaction evaluations of the robot's conversational qualities. Table~\ref{tab:rq3_descriptive} reports the mean, standard deviation, median, and interquartile range for naturalness, human-likeness, and interestingness. On average, participants rated the conversation as moderately natural and human-like, with interestingness receiving the highest ratings among the three qualities. These descriptive results provide an overview of participants' immediate impressions of the robot's dialogue behavior and serve as the basis of the correlation and regression analyses presented in the following subsections.

\begin{table}[h]
    \begin{tabular*}{\textwidth}{@{\extracolsep\fill}lccc}
    \toprule
    & Natural & Human-like & Interesting \\
    \midrule
    Mean   & 3.029 & 2.721 & 4.029 \\
    SD     & 0.939 & 0.921 & 1.014 \\
    Median & 3.0   & 3.0   & 4.0   \\
    IQR    & 1.25  & 1.0   & 1.0   \\
    \botrule
    \end{tabular*}
    \caption{Descriptive statistics for perceived conversational quality ratings after the verbal interaction. Values represent mean, standard deviation (SD), median, and interquartile range (IQR).}
    \label{tab:rq3_descriptive}
\end{table}

\subsubsection{Correlation Analysis}
\label{sec:4.3.2}
\begin{figure}[h]
    \centering
    \includegraphics[width=\linewidth]{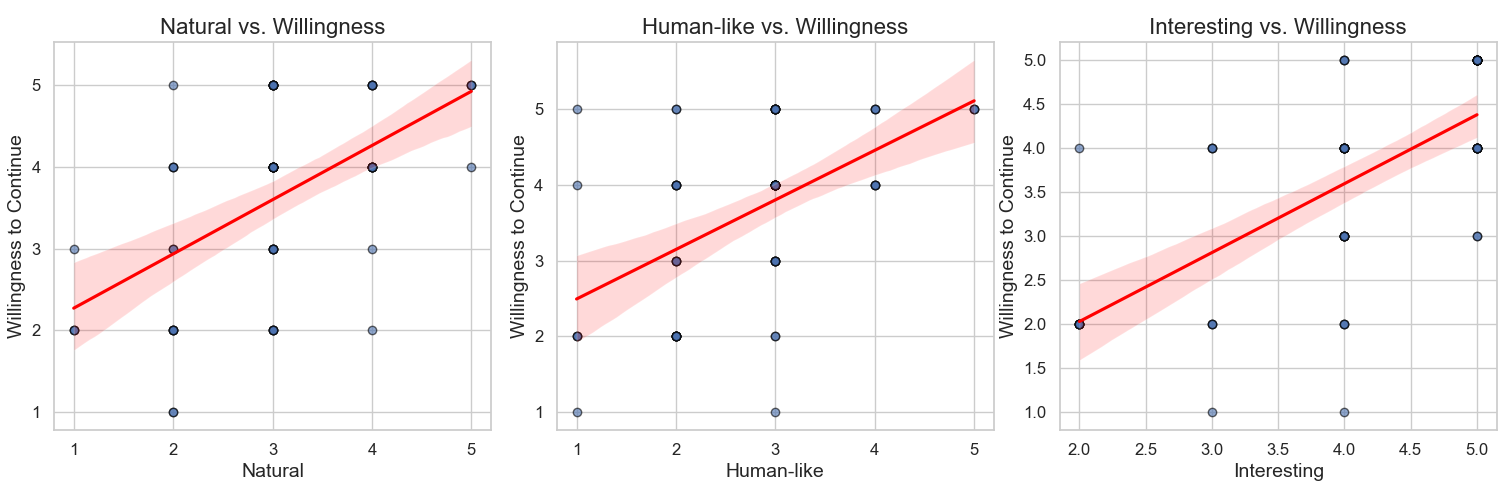}
    \caption{Scatter plots illustrating the relationships between conversational quality ratings and participants' intention to re-engage with the robot. Each plot displays the fitted linear regression line with 95\% confidence intervals (shaded areas), along with individual participant ratings for naturalness, human-likeness, and interestingness.}
    \label{fig:rq3_scatter}
\end{figure}

We examined the bivariate relationships between participants' evaluations of the robot's conversational qualities, including naturalness, human-likeness, and interestingness, and their intention to re-engage with the robot. Pearson correlation coefficients, corresponding $p$-values, and 95\% confidence intervals are reported in Table~\ref{tab:rq3_corr}. Figure~\ref{fig:rq3_scatter} presents scatter plots with fitted regression lines and 95\% confidence bands.

\begin{table}[h]
    \begin{tabular*}{\textwidth}{@{\extracolsep\fill}lccc}
    \toprule
    Predictor & Pearson $r$ & $p$ & 95\% CI \\
    \midrule
    Natural       & 0.535 & 0.000 & [0.340, 0.686] \\
    Human-like    & 0.518 & 0.000 & [0.319, 0.673] \\
    Interesting   & 0.682 & 0.000 & [0.530, 0.792] \\
    \botrule
    \end{tabular*}
    \caption{Pearson correlations between conversational quality ratings (naturalness, human-likeness, and interestingness) and participants' intention to re-engage with the robot. Reported values include correlation coefficients (Pearson $r$), $p$-values, and 95\% confidence intervals (CI).}
    \label{tab:rq3_corr}
\end{table}

All three conversational qualities were positively and significantly associated with intention to re-engage. Naturalness showed a moderate positive correlation with re-engagement intention ($r=0.535$, $p<.001$), as did human-likeness ($r=0.518$, $p<.001$). Interestingness demonstrated the strongest association with intention to re-engage ($r=0.682$, $p<.001$), with a confidence interval entirely above zero, and its magnitude was substantially higher than the other two predictors.

These results indicate that higher ratings of conversational quality were consistently associated with stronger intention to re-engage with the robot, with interestingness showing the most substantial relationship among the three conversational attributes.

\subsubsection{Multiple Regression Model}
\begin{figure}[h]
    \centering
    \includegraphics[width=0.9\linewidth]{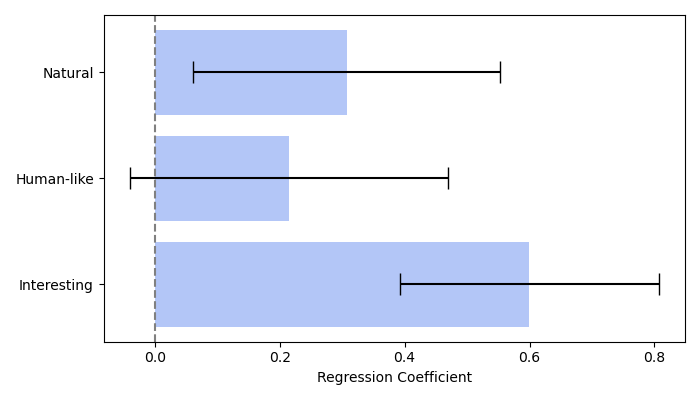}
    \caption{Coefficient plot showing the unique effects of conversational qualities on participants' intention to re-engage, based on a multiple linear regression model. Each bar represents the unstandardized regression coefficient, and the horizontal lines indicate the 95\% confidence intervals.}
    \label{fig:rq3_coefplot}
\end{figure}

To assess the unique contribution of each conversational quality to participants' intention to re-engage with the robot, we conducted a multiple linear regression including naturalness, human-likeness, and interestingness as simultaneous predictors. Figure~\ref{fig:rq3_coefplot} presents the unstandardized regression coefficients with 95\% confidence intervals, and Table~\ref{tab:rq3_regression} reports the full model estimates.

\begin{table}[h]
    \begin{tabular*}{\textwidth}{@{\extracolsep\fill}lcccccc}
    \toprule
    Variable & coef. & std err & $t$ & $p$ & std.\ coef. & 95\% CI \\
    \midrule
    Natural     & 0.307 & 0.123 & 2.492 & 0.015 & 0.248 & [0.061, 0.553] \\
    Human-like  & 0.215 & 0.128 & 1.689 & 0.096 & 0.170 & [-0.039, 0.470] \\
    Interesting & 0.599 & 0.104 & 5.774 & 0.000 & 0.522 & [0.392, 0.807] \\
    \botrule
    \end{tabular*}
    \caption{Multiple linear regression results predicting intention to re-engage from conversational quality variables. Unstandardized coefficients (coef.), standard errors (std err), $t$-values, $p$-values, standardized coefficients (std. coef.), and 95\% confidence intervals (CI) are reported.}
    \label{tab:rq3_regression}
\end{table}

Interestingness emerged as a significant positive predictor of re-engagement intention ($B=0.599$, $p<.001$), with a standardized coefficient of $0.552$ and a confidence interval that did not include zero. Naturalness also showed a statistically significant positive association with re-engagement intention ($B=0.307$, $p=.015$), though with a smaller standardized effect size ($0.248$). Human-likeness did not reach statistical significance in the model ($B=0.215$, $p=.096$), and its confidence interval included zero.

These findings indicate that, when considered together, interestingness and naturalness independently predict participants' intention to re-engage with the robot, with interestingness showing the strongest unique contribution. Human-likeness, although positively correlated with re-engagement intention (see Section~\ref{sec:4.3.2}), did not account for unique variance in the regression model once the other conversational qualities were included.

\section{Discussion}
\label{sec:5}
This study examined how users' perceptions of a humanoid robot evolve during a one-on-one verbal interaction and how these perceptions related to their intention to re-engage. Across three research questions, the findings highlight two central patterns. First, participants' perceived robot characteristics, including pleasantness, creepiness, and approachability, showed clear pre-post shifts following the interaction, indicating that even a brief verbal exchange can influence how users evaluate a robot's social qualities. Second, changes in these perceived characteristics did not significantly predict intention to re-engage, whereas participants' evaluations of the robot's conversational qualities, particularly interestingness and naturalness, were strongly associated with users' intention to re-engage with a robot. Together, these results suggest that while verbal interaction can shape affective impressions of the robot, users' re-engagement intentions are guided more directly by the perceived quality of the robot's conversational behavior.

\subsection{Changes in Perceived Robot Characteristics (RQ1)}
The results showed that participants' perceptions of the robot shifted significantly following the verbal interaction. Pleasantness and approachability increased, while creepiness decreased, reflecting an overall movement toward more positive evaluations of the robot's social qualities. These results echo prior studies showing that the presence and modality of verbal communication, such as human-like voice, expressive prosody, or emotional speech, can significantly influence users' affective impressions of social robots~\cite{kluber2025affect}. A brief conversational exchange may help clarify the robot's intentions, reduce uncertainty, and facilitate more favorable interpretations of its behavior and personality.

Although the present study focused exclusively on verbal interaction, the observed changes underscore the dynamic nature of users' perceptions in HRI. Perceived robot characteristics are not fixed judgments but can evolve quickly in response to the interactional context, which has implications for how robots are introduced and how initial impressions are formed during real-world deployment.

\subsection{The Role of Perceived Robot Characteristics in Engagement Intention (RQ2)}
Although participants' perceptions of the robot became more positive following the interaction, changes in pleasantness, creepiness, and approachability did not significantly predict their intention to re-engage.
This suggests that perceptual shifts alone may not be sufficient to drive engagement decisions. Several points from prior research help contextualize and explain this pattern.

First, in the broader literature of human-agent and human-robot interaction, engagement is recognized as a multifaceted concept, often comprising affective, cognitive, and behavioral components rather than a single evaluative dimension~\cite{oertel2020engagement}. Sorrentino et al.~\cite{sorrentino2024definition} note that engagement assessments typically rely on observable behaviors or self-reported global evaluations, rather than on short-term changes in affective impressions. This aligns with our finding that although perceptions became more positive, these shifts did not correspond to re-engagement intention.

Second, studies often highlight the influence of a robot's appearance or anthropomorphic features on initial impressions, their role in long-term or repeated engagement, however, is less clear. Prior work shows that such cues matter early in an interaction but become less influential once users begin assessing the robot's behavioral performance, particularly its communicative behaviors~\cite{ahmad2025understanding}. In this context, our results suggest that once the interaction unfolds, conversational behaviors may outweigh perceptual impressions in shaping users' intention to re-engage.

In sum, existing literature supports the interpretation that perceptual changes alone may not suffice to predict re-engagement intentions. The present findings reinforce the importance of focusing on interactive and conversational behaviors rather than static robot features when designing systems intended to support sustained engagement.

\subsection{The Role of Conversational Qualities in Engagement Intention (RQ3)}
Our findings show that participants' perceptions of the robot's conversational interestingness and naturalness were both strongly associated with and significant unique predictors of, their intention to re-engage with the robot. This suggests that, in the context of HRI, conversations that are perceived as engaging and natural are critical to fostering users' motivation for sustained engagement. In contrast, human-likeness, while positively associated at the bivariate level, did not emerge as a significant independent predictor in the multivariate analysis.

These results align with previous studies of user engagement in HRI, which emphasize the importance of perceived quality and authenticity in fostering user engagement~\cite{cai2024communication,park2023examining}. Building on this perspective, our results indicate that, within the context of conversational interaction, perceived interestingness and naturalness of dialogue are more influential for sustained engagement than the perceived human-likeness of the robot's conversational style. When conversations with robots are felt to be interesting and flow naturally, users may become more emotionally and cognitively involved, thereby increasing their desire to continue the interaction~\cite{irfan2024human,chen2025impact}. 

The findings that human-likeness lost significance in the multivariate model may indicate that, although users appreciate human-like qualities, these are not sufficient on their own to sustain engagement when more direct conversational attributes, such as interesting content and natural flow, are present. This supports the argument that overemphasis on anthropomorphic cues in conversational style may not add value beyond what is achieved through authentic, high-quality verbal interaction~\cite{cai2024communication}.

From a design perspective, these results highlight the need for prioritizing interestingness and naturalness in robot dialogue systems. Designers and developers should focus on creating conversations that are contextually relevant, stimulating, and responsive to user input, as well as ensuring smooth, coherent turn-taking and timing. While maintaining some human-like qualities can be beneficial, the primary goal should be to deliver interactions that are perceived as engaging and authentic.

Recommendations for developers include:
\begin{itemize}
    \item investing in dialogue management systems that can sustain user interest and adapt to conversational context.
    \item incorporating personalization and content variability to keep conversations fresh and interesting.
    \item using human-like cues judiciously, ensuring that anthropomorphism supports rather than substitutes for engaging and natural dialogue.
\end{itemize}

In summary, fostering conversations that are interesting and natural may be the most effective way to promote sustained user engagement with social robots, potentially offering greater benefits than simply making the robot’s conversational style more human-like.

\subsection{Limitations and Directions of Future Research}
While this study provides important insights into the affective and conversational factors influencing engagement with social robots, several limitations should be considered when interpreting the findings.

First, the generalizability of these results may be limited by the specific context in which the study was conducted. The interaction scenario, the particular robot used, and the characteristics of the participant sample (e.g., students, specific age group, cultural background) may not reflect the experiences of broader or more diverse populations. Future research should investigate the robustness and generalizability of these findings by testing them with diverse user populations, robot platforms, and interaction contexts.

Second, although the study examined key affective and conversational predictors, the results revealed a notable discrepancy between bivariate and multivariate analyses. Specifically, no single change in perceived robot characteristics emerged as a dominant predictor of sustained engagement; instead, each appeared to contribute only modestly and somewhat independently. Furthermore, the absence of unique significant predictors may reflect limitations in sample size or measurement precision, underscoring the need for larger and more robust studies to clarify these relationships.

Third, the current investigation focused on a limited set of affective and conversational predictors. However, engagement in HRI is likely to be shaped by a broader range of factors, such as empathy, adaptability, or social characteristics~\cite{oertel2020engagement,garcia2023empathy,salam2016fully}. Exploring these qualities in future research may reveal additional pathways to sustained engagement.

Finally, this study assessed users' responses in a single, relatively short-term interaction. It is unclear whether the observed changes in perceived robot characteristics and the influence of conversational qualities on users' behavioral intention would persist over multiple or extended interactions. Longitudinal research is needed to determine the stability and evolution of these effects over time.

Addressing these limitations and research directions will contribute to a more comprehensive understanding of the mechanisms underlying engagement in HRI and inform the design of more effective, socially and affectively responsive robotic systems.

\section{Conclusion}
\label{sec:6}
This study provides empirical evidence that both affective evaluations and conversational qualities influence users’ intention to re-engage with humanoid social robots. While verbal communication by the robot led to more positive perceptions of the robot's affective characteristics, including increased pleasantness and approachability and reduced creepiness, these changes alone were not strong or unique predictors of users’ intention to re-engage. Instead, it was the perceived interestingness and naturalness of the robot’s conversation that emerged as robust, independent predictors of users' intention to re-engage.

These findings suggest that, for sustaining user engagement in HRI, the quality and meaningfulness of the verbal exchange play a more central role than affective perceptions alone or the perceived human-likeness of the robot's conversational style. While anthropomorphic features may shape initial impressions, our results suggest that authentic, contextually relevant, and smoothly flowing conversations are ultimately more influential in motivating users to return for subsequent interactions.

For designers and developers, this highlights the value of investing in advanced dialogue systems capable of maintaining user interest, adapting to conversational context, and supporting engaging, natural interactions. Anthropomorphic features, such as human-like language style or social cues, should be viewed as complementary enhancements rather than substitutes for strong conversational design.

\section*{Statements and Declarations}

\subsection*{Data Availability Statement}
The data that support the findings of this study are not publicly available due to participant privacy but are available from the corresponding author upon reasonable request.

\subsection*{Competing Interests and Funding}
The authors have no competing interests to declare that are relevant to the content of this article. No funding was received for conducting this study.

\subsection*{Ethics Approval}
All procedures performed in studies involving human participants were in accordance with the ethical standards of the University of Geneva's Committee for Ethical Research and with the 1964 Helsinki Declaration and its later amendments or comparable ethical standards. The study was approved by the University of Geneva's Committee for Ethical Research (CUREG-2024-10-109).

\subsection*{Author Contributions}
Conceptualization: Hangyeol Kang, Nadia Thalmann; Methodology: Hangyeol Kang, Thiago Freitas, Maher Ben Moussa; Formal analysis and investigation: Hangyeol Kang, Thiago Freitas; Writing - original draft preparation: Hangyeol Kang; Writing - review and editing: Hangyeol Kang, Thiago Freitas, Maher Ben Moussa, Nadia Thalmann; Resources: Nadia Thalmann; Supervision: Nadia Thalmann.



\bibliography{sn-bibliography}

\end{document}